\documentclass[amsmath,amssymb,aps,prd,floatfix,12pt,preprintnumbers]{revtex4-1}

\usepackage{bm}

\usepackage{amsfonts,amsbsy,latexsym,amssymb,amscd,amstext}
\usepackage{epsfig}
\usepackage{graphicx}
\usepackage{subfigure}
\newcommand{\be}{\begin{equation}}
\newcommand{\ee}{\end{equation}}
\newcommand{\ba}{\begin{array}}
\newcommand{\ea}{\end{array}}
\newcommand{\baa}{\begin{array}}
\newcommand{\eaa}{\end{array}}
\newcommand{\bea}{\begin{eqnarray}}
\newcommand{\eea}{\end{eqnarray}}

\begin{document}
\preprint{IFT-UAM/CSIC-12-32, FTUAM-12-89}
\title{  Real-time Quantum evolution  in the  Classical approximation and beyond}

\author{ Antonio Gonz\'alez-Arroyo $^{a,b}$ and Ferm\'{\i}n Nuevo $^{a}$}
\affiliation{$^a$ Instituto de F\'{\i}sica Te\'orica UAM/CSIC \\
$^b$ Departamento de F\'{\i}sica Te\'orica, C-15 \\
Universidad Aut\'onoma de Madrid, E-28049--Madrid, Spain}
\email{antonio.gonzalez-arroyo@uam.es, fermin.nuevo@uam.es}

\begin{abstract}
With the goal in mind of deriving a method to compute quantum
corrections for the real-time evolution in quantum field theory, 
we analyze the problem from the perspective of the Wigner function. 
We argue that this provides the most natural way to justify and 
extend the classical approximation. A simple proposal is presented 
that can allow to give systematic quantum corrections to the evolution
of expectation values and/or an estimate of the errors committed when
using the classical approximation. The method is applied to the case of a 
few degrees of freedom and compared with other methods and with the
exact quantum results. An analysis of the dependence of the numerical 
effort involved as a function of the number of variables is given, 
which allow us to be optimistic about its applicability in a quantum
field theoretical context.
\end{abstract}




\date{\today}


\maketitle



{\vskip 1cm}

\section{Introduction}
\label{s.intro}
Quantum Phenomena underlie most of Modern Physics. Alongside it brings
in the necessity of dealing with complex functions and interference
phenomena. In particular, determining the time evolution of a quantum 
system is relevant for many areas of Physics. When a large number of 
degrees of  freedom is involved, numerical methods based on the integration of 
the Schroedinger equation fail. This, however, is the typical situation 
in Quantum Field Theory (QFT). \nocite{*}

When computing  expectation values of operators in the vacuum or in
the  equilibrium state, one can use the path-integral approach. 
Furthermore, 
in the computation of  the Wick-rotated Green functions of the theory, 
the complex quantum weight of each trajectory
is  transformed into a positive-definite  probability weight. This
allows the use of efficient  standard importance sampling
techniques, such as the Metropolis algorithm or other Monte Carlo techniques.
This information allows the extraction of the spectrum and other properties of the
theory. The same applies when studying Quantum Field Theory at
equilibrium. However, even in this situation, there are important exceptions
in which the  weights are not positive definite and the standard 
Monte Carlo methods fail. This is often referred as {\em the sign
problem}. One example occurs in certain Quantum 
Field theories, as QCD, at finite chemical potential. Many methods have been
devised to obtain relevant information in this situation, meeting
partial success~\cite{deForcrand:2009ys}. However, it is generally accepted that, despite the
efforts, no fully satisfactory solution has been found. This is
particularly unwelcome, since full-proof predictions in certain areas
of Physics,  which are of great relevance and timeliness (such as
Heavy Ion collisions) are lacking. 

When studying the quantum evolution away from equilibrium or from a  initial
state which is not an eigenstate of the Hamiltonian the situation is
even more severe. In this context, path integral methods for
time-dependent expectation values  follow from the
Schwinger-Keldysh formalism~\cite{Schwinger:1960qe}\cite{Keldysh:1964ud}. This allows systematic 
perturbative calculations. However, non-perturbative
phenomena are often crucial and we lack an efficient numerical 
computational method to deal with this situation (for a review see
Ref.~\cite{Berges:2004yj}).

This problem  arises in different areas of Physics such as
Nuclear Physics, Quantum Chemistry, Quantum Optics, etc.
One of these areas is Cosmology, which actually  triggered the interest of the
present authors in the problem. Quantum fluctuations play a role at different 
instances in the early Universe. 
One such case is at the inflationary era, by generating the density fluctuations which 
act as sources for the anisotropies of the cosmic background radiation and the 
formation of structures. Many authors argued that the fluctuations
develop from quantum  to classical, and can be treated as classical at
late times~\cite{Guth:1985ya}-\cite{Koksma:2011dy}. Another interesting epoch which depends 
crucially on the understanding of the quantum evolution of a quantum field theory,
is that of  preheating and reheating after
inflation~\cite{Kofman:1994rk}-\cite{Greene:1997fu}.  Properties of the
present Universe, such as baryon number density, gravitational waves or
magnetic field remnants, might depend upon this dynamics. Obtaining
reliable estimates demand an appropriate treatment of the quantum
field theory evolution from an initial state after inflation to the
fully thermalized reheated Universe. To estimate these effects, several
authors~\cite{Khlebnikov:1996mc}-\cite{DiazGil:2008tf} have employed the
so-called {\em classical approximation} (For a somewhat different
context see also Refs.~\cite{Aarts:2001yn}-\cite{Aarts:2001yx}). This consists on treating  certain modes
of the  quantum fields  as random  fields with deterministic dynamics given 
by the classical equations of 
motion of the system.  The randomness is imprinted in the initial conditions 
of the fields, often determined by a particularly simple initial quantum state.
The authors who employ the classical approximation, often present arguments
to  determine which modes can indeed be treated as classical and those
that cannot, leading to different types of initial conditions in both
cases. The last point is crucial since in Quantum Field Theory there
are ultraviolet divergences. An appropriate treatment has to deal
with them through Renormalization. 

The present work originated with the goal of determining the region of 
validity of the classical approximation, estimating 
the size of the errors induced by it, and hopefully even obtaining a way
to go beyond it, by  incorporating quantum corrections to the results. For
that purpose we have started by analyzing simple quantum systems with very few degrees
of freedom whose quantum evolution can be obtained by numerical
integration of the Schroedinger equation. The initial conditions and 
the type of Hamiltonians considered  are inspired by the field
theoretical  and cosmological applications. Thus, we focus  upon
quartic interactions and gaussian or thermal initial conditions. 

In all cases we have compared the results obtained by the classical
approximation with the exact quantum evolution. Furthermore, we have 
also studied the results obtained with different proposals for 
incorporating quantum corrections, limiting ourselves to those that
have any hope of being applicable to the quantum field  theoretical
case. In particular, a very interesting technique is the two-particle
irreducible effective action supplemented with a certain truncation 
of the number of diagrams involved (2PI
method)~\cite{PhysRev.127.1391}-~\cite{Cornwall:1974vz}. 
This truncation can be based on the loop expansion or on the 1/N
expansion~\cite{Berges:2001fi}. The latter behaves in a
more  stable fashion and has been used in our analysis. We remark that 
the traditional Hartree method can be considered a particular case of
the 2PI method~\cite{PhysRevD.36.3114}-\cite{PhysRevD.65.045012}.  In any case the method is limited to the determination 
of the quantum evolution of certain observables, such as the 2-point 
correlation function of the system (see Ref.~\cite{Berges:2004yj} for a more
complete list of early references on the subject). 

Another technique which has been recently applied in the context of Quantum Field Theory is the 
{\em complex Langevin method}~\cite{Klauder:1983zm}-\cite{Berges:2005yt}. This is based on 
complexifying the fields and studying the dynamics of the field 
trajectories induced by a Langevin equation in an additional 
Langevin-time variable with a purely imaginary drift term. Instabilities are 
often found in the numerical integration of this equation, although
authors have given several recipes to avoid them~\cite{Aarts:2009dg}. Furthermore,  
good results have been reported in certain
cases~\cite{Aarts:2009uq}-\cite{Aarts:2011ax}, so that  we thought it
was very interesting  to apply it to our examples. Unfortunately, we seem to be
in a situation in which convergence to the right solution does not
apply. This question deserves future study. 

In parallel to the tests explained before, we present a method to 
quantify and incorporate quantum effects based upon the {\em Wigner
function}~\cite{Wigner:1932eb}. This  is a pseudo-distribution
function, whose expectation values give us the matrix elements  of 
Weyl-ordered products of operators in the quantum state of the system.
The function is real, but not positive definite (see
Ref.~\cite{Hillery:1983ms} for an account of all its properties). The Wigner function 
satisfies an evolution equation~\cite{Moyal:1949sk} in time, which determines the
time-dependence of all these expectation values. One of the advantages 
of this method, is that it is particularly simple to see what is the meaning
of the classical evolution and what is the nature of quantum corrections. 
This will be explained in the next section. This observation is not
novel, and has led  different researchers in different fields to focus
on the Wigner function and its evolution equation when  attempting to 
describe quantum evolution~\cite{Remler:1975fm}-\cite{Kunihiro:2008gv}.
One example, is in Nuclear Physics where 
several authors~\cite{Bonasera:1993zz}\cite{John:2007zz}  have developed methodologies which
are very similar in spirit to our goal (see also \cite{Remler_1987}). Unfortunately, the detailed 
techniques seem hard to extend to a large number of degrees and thus to 
Quantum field theory. In our particular proposal we have dedicated
some time to study the way in which the numerical effort involved
scales with the number of degrees of freedom. A power-like growth is
acceptable even if it involves an enormous computational effort.
Experience teaches us that the development of computer technology and
algorithms will diminish the load in due time. An exponential growth is 
a killer. Our results presented below are promising and seem to allow 
the computation of quantum corrections in typical situations relevant
for cosmological applications. This will be addressed in a future paper 
of the present authors. The present paper is to be considered  a pilot study, 
in which we have the advantage of knowing what the exact quantum
result is. The full field theoretical case demands  a much 
higher computational cost, in addition with the necessity of 
dealing with issues as  renormalization, as  mentioned previously. 

For a more complete account of the literature we should mention 
the work of~\cite{Mrowczynski:1994nf}, in which the authors advocate the use of the 
Wigner function evolution equation in Quantum Field Theory and give
ideas  on possible techniques to obtain an explicit calculation of the 
quantum corrections in this setting. 

We conclude this section by describing the lay-out of the paper. 
In the next section we review the definition of the Wigner function
and derive its time-evolution equation. We also explain the meaning 
of the classical approximation in this context. The following section
explains how, in trying to derive the quantum Liouville equation as a 
Fokker-Planck equation associated to a Langevin process, the
non-positive definite character leads to pathological properties of the 
stochastic force. With this idea in mind, in Section~\ref{smethod} we
present a method to  relate the equation to a regular
markovian process for which standard sample methods can be applied.
This together with a certain coarse-graining approximation allows to
set up a procedure to calculate systematic quantum corrections to the
evolution in powers of $\hbar^2$. In the following section, we study 
several simple cases for which we  analyze the accuracy of the classical
approximation, the truncated 2PI method and our proposal of the
previous section. Special attention is paid to estimate the 
capacity of the method to deal with discretized lattice approximations
to quantum field theory. In the concluding section, we summarize our
results and discuss  the advantages and limitations of our 
proposal.

\section{The Wigner function evolution equation}

In Quantum Mechanics the expectation values of Weyl-ordered products of operators
can be computed in terms of the Wigner function $W(x,p,t)$ as
follows\cite{Wigner:1932eb}\cite{Hillery:1983ms}:
\be
\langle \Psi | f_W(Q,P) | \Psi \rangle =\int \frac{ dx\, dp}{\hbar}
f(x,p) \, W(x,p,t)
\ee
where $f_W(Q,P)$ means a Weyl-ordered product of position and momentum
operators, whose classical limit is $f(x,p)$.  
For a pure state, given the wave function $\Psi(x,t)$, the expression of the Wigner function is
\be
\label{wig_wavef}
W(x,p,t) = \int \frac{dy}{2 \pi} \, \Psi^*(x+y/2,t) \Psi(x-y/2,t)
 \, e^{i py/\hbar} 
\ee 
This can be extended to mixed states associated to a density matrix
$\mathbf{\rho}(t)$ as follows:
\be
W(x,p,t) = \int \frac{dy}{2 \pi} \, \langle x-y/2 | \mathbf{\rho}(t) |
x+y/2 \rangle
 \, e^{i py/\hbar} 
 \ee
In  both cases the Wigner function is real but not necessarily
positive definite.

The Wigner function satisfies an evolution equation in time which
depends on the form of the potential. Here we will
derive it for the case of Hamiltonian of the form
\be
H=\frac{p^2}{2 m} + V(x) 
\ee
The corresponding Schroedinger equation satisfied by the wave function
is 
\be
i \partial_0 \Psi(x,t) = -\frac{\hbar}{2
m}\frac{\partial^2}{\partial x^2} \Psi(x,t) + \frac{V(x)}{\hbar} \Psi(x,t) 
\ee
From this equation, using the relation Eq.~\ref{wig_wavef}, one can derive the equation followed by the Wigner
function:
\begin{eqnarray}
\nonumber
\partial_0 W(x,p, t) &= \frac{i \hbar}{2 m} \int \frac{dy}{2 \pi}
\,e^{i py/\hbar} \left(\Psi^*(x+\frac{y}{2},t)
\frac{\partial^2\Psi(x-\frac{y}{2},t)}{\partial x^2} - \frac{\partial^2
\Psi^*(x+\frac{y}{2},t)}{\partial x^2} \Psi(x-\frac{y}{2},t)\right) + \\
&+ \frac{i}{\hbar} \int \frac{dy}{2 \pi} \, \Psi^*(x+\frac{y}{2},t) \,
  \Psi(x-\frac{y}{2},t) \, e^{i py/\hbar}
(V(x+\frac{y}{2})-V(x-\frac{y}{2})) \:\:\:
\end{eqnarray}
One can transform the right hand side in an obvious way and obtain 
\begin{equation}
\nonumber
 \int \frac{dy}{2 \pi} \,e^{i py/\hbar} \left(\frac{-i \hbar}{ m}
\frac{\partial^2}{\partial x \partial y} +\frac{i}{\hbar}
\left(V(x+\frac{y}{2})-V(x-\frac{y}{2})\right)\right)
\Psi^*(x+\frac{y}{2},t) \, \Psi(x-\frac{y}{2},t)
\end{equation}
Integrating by parts one can substitute $\frac{\partial}{\partial y}$ by
$-ip/\hbar$. On the other hand, the factors of y can be replaced by 
$-i\hbar \frac{\partial}{\partial p}$. We end up with the equation
\begin{equation}
\partial_0 W(x,p, t) = -\frac{p}{m} \frac{\partial W(x,p, t)}{\partial
x } +\frac{i}{\hbar}  \left(V(x-\frac{i\hbar}{2}\frac{\partial}{\partial
p})-V(x+\frac{i\hbar}{2}\frac{\partial}{\partial p})\right) W(x,p,t)
\end{equation}
This equation is well-known~\cite{Moyal:1949sk} and receives several names in the
literature: Moyal equation or quantum Liouville equation. 

Let us now restrict  to a potential of the form 
\be
\label{potential}
V(x)= \frac{\mu^2}{2} x^2 + \frac{\lambda}{4!} x^4 
\ee
The operator involving $V$ can be expanded to give 
\be 
V'(x) \frac{\partial}{\partial p} -\frac{\hbar^2}{24}
V'''(x)\frac{\partial^3}{\partial p^3} 
\ee

Then, we are finally led to an equation of the form: 
\begin{equation}
\label{main_eq}
\partial_0 W(x,p, t) = -\frac{p}{m} \hspace{1pt} \frac{\partial W(x,p, t)}{\partial
x } + V'(x) \hspace{1pt} \frac{\partial W(x,p,t)}{\partial p} - \frac{\lambda
\hbar^2 x}{24} \hspace{1pt} \frac{\partial^3 W(x,p, t)}{\partial p^3}
\end{equation}
Notice that the first two terms do not contain $\hbar$. As a matter of
fact, if we neglect the last term, the solution to this partial
differential equation is very simple. It is given by 
$$ W_0(x_0(x,p,t), p_0(x,p,t))$$ 
where the functions $x_0$ and $p_0$ are obtained by running back in
time to time zero the classical equations of motion from a point
$(x,p)$ in phase-space at time t.   This
is the so-called  {\em classical approximation} to the quantum evolution. 

The last term contains all quantum effects and has dramatic consequences.
In particular, the Wigner function is not guaranteed to remain
positive at all times. Thus, computing expectation values with the Wigner
function can be very unstable numerically, because it comes from a
cancellation of both positive and negative terms which might be much
larger than the overall sum. This is a typical {\em sign problem}, which
might render difficult to compute quantum expectation values by 
probability methods. However, if we start at $t=0$ from a positive Wigner
function it might take some time until the negative part contributes
sizably, and expectation values can be determined with reasonable 
accuracy.

The size of the last term is, in principle,  small in macroscopic terms,
being proportional to  $\hbar^2$. However, this depends very much on the
size of the third derivative of the Wigner function. This a time-dependent 
function, but
it is clear that the initial distribution has an important effect on
the accuracy of the classical approximation at initial times. 
This can be tested in our  particular cases, since one can numerically integrate 
the Schroedinger equation. 

If one focuses upon expectation values, the size of quantum effects
and the errors committed by numerical integration of the Moyal equation 
can be quite different. It is to be expected that the accuracy  of  
expectation values is better for operators involving $Q$ alone, than for those
involving $P$.

Let us restrict to a gaussian  initial distribution, which appears
naturally in many applications. Assuming for simplicity factorization
of the distribution in $x$ and $p$, the initial Wigner function takes the form:
\begin{equation}
W(x,p,t=0)= \frac{\hbar}{2 \pi \sigma_x \sigma_p} \exp\{-\frac{p^2}{2
\sigma_p^2} -\frac{x^2}{2
\sigma_x^2} \}
\end{equation}
For a pure state,  the two standard deviations should  be
related as follows:
$$ \sigma_x \sigma_p = \frac{\hbar}{2} $$
For a mixed state this condition is relaxed. For example, for the density matrix of
a harmonic oscillator at thermal equilibrium, this product 
is given by 
$$ \sigma_x \sigma_p = \frac{\hbar}{2 \tanh(\hbar \omega \beta/2)} $$
This interpolates between $\hbar/2$ at low temperatures and $kT/\omega$ at high
temperatures.

Sticking to the pure state case, and given the scales involved in the
problem, one can form dimensionless
quantities which control the relative importance of quantum effects. 
The first one $r$ is the usual one formed by taking the ratio 
of a classical quantity with the dimensions of action, divided by
$\hbar$. In our present case, this quantity is 
\be
r=\frac{\sqrt{m}\mu^3}{\lambda \sigma_x \sigma_p}
\ee
One expects smaller quantum effects for large values of $r$. 
However, there is another combination which seems more directly related 
to the size of the quantum term in the equation for the Wigner
function. This is given by 
\be 
s=\frac{\sigma_p \mu}{\sigma_x^3 \lambda \sqrt{m}}
\ee

\begin{figure}
\begin{center}
\includegraphics[angle=-90, width=0.7\textwidth]{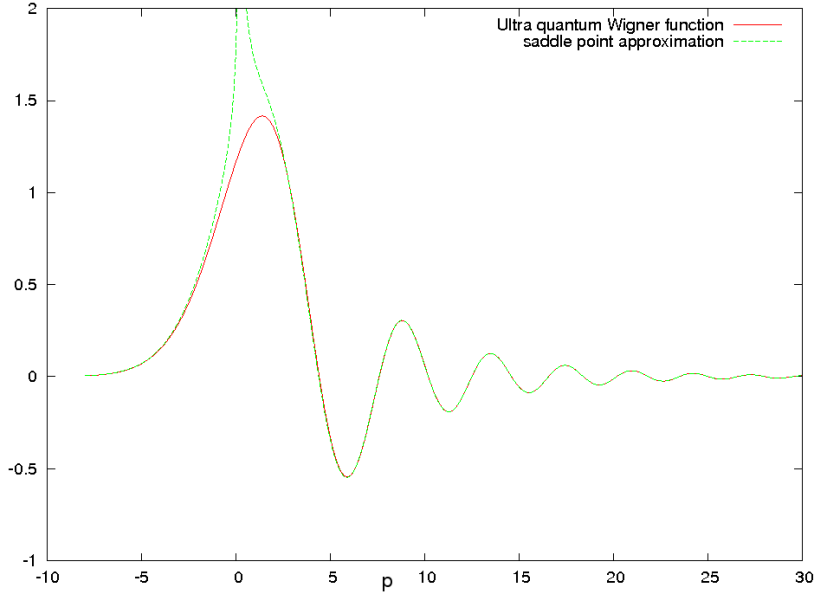}
\caption{The Wigner function $W_{\mathrm{uq}}$ in the ultraquantum
limit, for fixed values of $x$ and $t$.}
\label{fig1}
\end{center}
\end{figure}

To get some insight into the structure of the Wigner function, we can
study  certain limits. For example, one can consider 
an {\em ultra-quantum} limit, in which  we  neglect the classical
$\hbar$-independent terms in the equation satisfied by the 
Wigner function. The equation can be integrated exactly in this case, 
and the result is a one-dimensional integral:
\be
W_{\mathrm{uq}}(x,p,t)= \frac{\hbar}{(2\pi)^{3/2}\sigma_x} \hspace{4pt} e^{-x^2/(2 \sigma_x^2)} \int dz \exp\{ipz -i
\lambda Q z^3  - \sigma_p^2 y^2/2\}
\ee
where $Q\equiv -\lambda \hbar^2 x t/24$. This can be related to Airy
functions. The shape of this function is displayed in Fig.\ref{fig1}
for $Q= 2$ and $\sigma_p^2=2$.  Notice
the damped oscillations for positive $p$. It is clear that the Wigner
function becomes negative in some regions, but the total integral
is finite and positive. As a matter of fact, it is
quite easy to understand this oscillatory pattern and its dependence
on $x$ and $t$, by evaluating the integral in the saddle point
approximation. The result is also plotted in Fig.\ref{fig1}.
For small
values of $p$, the approximation breaks down as expected, but it
becomes quite precise for large values of $|p|$, which encompasses the
oscillatory region. Introducing $\kappa= 3Qp-\sigma_p^2/4$, the
approximation is different for positive and negative values of
$\kappa$. In the first case we have 
\be 
2 \sqrt{\pi} \kappa^{-1/4} \cos(2 \kappa^{3/2}/(27
Q^2)-\pi/4) \, \exp\{-p \sigma_p^2/(6Q)+ \sigma_p^3/(108 Q^2)\}
\ee
while for negative $\kappa$ we have 
\be
\sqrt{\pi} |\kappa|^{-1/4} \exp\{-2 |\kappa|^{3/2}/(27
Q^2) -p \sigma_p^2/(6Q)+ \sigma_p^3/(108 Q^2)\}
\ee
Notice that for large $p$ the argument of the cosine is proportional to 
$p^{3/2}/Q^{1/2}$, so that it broadens for large times. 
Despite the complicated behavior of the Wigner function in this
ultraquantum case, all expectation values of $f(x)$ are time
independent. 

One can go beyond this approximation by considering also the second
term on the right-hand side of the Wigner function equation. This
approximation is equivalent to the infinite mass limit
$m\longrightarrow \infty$. The new Wigner function is obtained from 
$W_{\mathrm{uq}}(x,p,t)$ by replacing $p$ by $p+V'(x)t$. Again, all
expectation values of $x$ are time independent.

\section*{Many variables}
The previous results generalize to more than one variable in a fairly
straightforward fashion. The equation satisfied by the Wigner function 
becomes: 
\be
\partial_0 W(\vec{x}, \vec{p},t)= -\sum_n \left(\frac{p_n}{m} \frac{\partial
W}{\partial
x_n } +\frac{\partial V}{\partial x_n}
\frac{\partial W}{\partial p_n}\right) -
\frac{\hbar^2}{24}  \sum_{n,m,r} \frac{\partial^3 V }{\partial x_n
\partial x_m \partial x_r} \, \frac{\partial^3 W  }{\partial p_n
\partial p_m \partial p_r}
\ee

Now we will consider two particularly important cases. The first one
is a O(N) symmetric potential:
\begin{equation}
V= \frac{\mu^2}{2} ||\vec{x}||^2 +\frac{\lambda}{8}
\left(||\vec{x}||^2\right)^2
\end{equation}
With this potential the last term on the right-hand side of the Wigner
function equation becomes
\begin{equation}
\frac{\hbar^2 \lambda}{8} \sum_n \sum_m x_n \frac{\partial^3 W}{\partial p_n
\partial p_m^2}
\end{equation}
An interesting situation occurs when the initial distribution is O(N)
invariant. The Wigner function at all times would only depend on
invariants ($A=\vec{x}\cdot\vec{p}$, $B=||\vec{p}||^2$, and
$C=||\vec{x}||^2$. The corresponding equation for $W(\vec{x},
\vec{p},t)=F(A,B,C,t)$ 
is given by 
\begin{eqnarray}
\nonumber
&\partial_0 F=(-\frac{B}{m}+\mu^2 C+\frac{\lambda}{2} C^2)
\frac{\partial F}{\partial A}- \frac{2 A}{m}
\frac{\partial F}{\partial C}+  2 A (\mu^2 +\frac{\lambda}{2}
C)\frac{\partial F}{\partial B}- \\
\nonumber
&\hspace{2.9cm}-\frac{\lambda \hbar^2}{8}\left( C^2\frac{\partial^3 F}{\partial A^3}+
6AC\frac{\partial^3 F}{\partial A^2\partial B}+4
(2A^2+BC)\frac{\partial^3 F}{\partial B^2\partial A} + 8 AB
\frac{\partial^3 F}{\partial B^3} \, + \right. \\
&\left. + \, (2N+4)C \frac{\partial^2 F}{\partial A\partial B} +(4N+8)A
\frac{\partial^2 F}{\partial B^2}\right) \hspace{1.7cm}
\end{eqnarray}
If the initial Wigner function has typical values of $A$, $B$ and $C$
proportional to $N$, as in the case of independent variables, and we
scale $\lambda$ to be proportional to $1/N$, the quantum evolution
preserves these properties. It is interesting to notice, that in this
case the quantum-term in the evolution of the Wigner function is
suppressed by one or two powers of $N$ in the denominator. This means
that in this particular large $N$ limit, the typical expansion
parameters for the quantum evolution is 
$$  \frac{\hbar^3
\lambda }{16 N \sigma_p^4}$$ 
This is consistent with  the conventional assertion  that the large N dynamics is
classical. Furthermore, notice that those terms containing third
derivatives are suppressed by two powers of $N$, instead of one. 
Thus, to leading order in $1/N$ the Wigner function satisfies a
simplified equation containing only second derivatives. After some work 
one can write this leading quantum term as: 
\begin{equation}
-\frac{\hbar^2N\lambda}{8(BC-A^2)} \sum_{n,m} x_m (Cp_n-Ax_n) \frac{\partial^2
F}{\partial p_n\partial p_m} 
\end{equation}

The second case which we want to consider is that in which the
coordinates are labeled by  $\vec{n}$, the points of a d-dimensional
hypercubic lattice $\Lambda$. The coordinates will be referred as 
 $\phi(\vec{n})$, and the corresponding Hamiltonian is given by
\be \label{ewigfields01}
H=\sum_{\vec{n}\in \Lambda} \left( \frac{\pi(\vec{n})^2}{2 m} +
\frac{1}{2a^2}\sum_\mu
(\phi(\vec{n}+\vec{\mu})-\phi(\vec{n}))^2+\frac{\mu^2}{2} \phi^2(\vec{n})
+\frac{\lambda}{24} \phi^4(\vec{n}) \right)
\ee
The main property of this family of Hamiltonians is its invariance
under the symmetry group of translations in d dimensions. 
Notice that this corresponds to the discretization of the Hamiltonian
of a $\lambda \phi^4$ scalar field theory in d-dimensions on a lattice of spacing
$a$. The quantity $\pi(\vec{n})$ is the conjugate momentum to
$\phi(\vec{n})$, satisfying canonical commutation relations among them. 
If we apply the general  formulas to derive the quantum Liouville
equation for the Wigner function in this case, we obtain that the 
 quantum term is given by:
\be \label{ewigfields02}
-\frac{\lambda \hbar^2}{24}  \sum_{\vec{n}\in \Lambda} \phi(\vec{n})
\frac{ \partial^3 W}{\partial \pi(\vec{n})^3  }
\ee
Formally, it is possible to take the continuum limit $a\longrightarrow
0$ and write down the equation satisfied by the Wigner function in the
case of quantum field theory. Here we will not be using it so we will
not write it down explicitly. The reader can consult Ref.~\cite{Mrowczynski:1994nf}
where it is spelled out.


\section{Langevin approach to  quantum evolution}
In this section we will try to generate the Wigner function quantum
evolution equation by means of a Langevin process. A typical Langevin 
dynamical process cannot do the job, since it preserves the positive
character of the probability density. Hence, we need a non-trivial
modification of the standard technique to apply to this case. In what
follows we will see how this comes out exactly.

Let us begin by writing the classical evolution equations with the
addition of a random force:
\begin{eqnarray} \label{emotlan1d}
&\dot{x}(t) = \frac{p(t)}{m} \:\:\:\:\:\:\:\:\:\:\:\:\:\:\:\:\:\:\:\:\:\:\:\:\:\:\: \\
&\dot{p}(t) = -V'(x(t)) + F(t)
\end{eqnarray}
Our goal will be to study the properties of the force $F(t)$ to recover
the equation for the Wigner function. In order to do so in a
simplified manner, let us discretize the time variable and write the 
equation relating $x'\equiv x(t+\delta t)$ and $p'\equiv p(t\ +\delta
t)$ to $x\equiv x(t)$ and
$p\equiv p(t)$:
\begin{eqnarray}
&x'= x+ \delta x = x+ \delta t \, \frac{p}{m} \,\:\:\:\:\:\:\:\:\:\:\:\:\:\:\:\:\:\:\:\:\:\: \\
&p' = p +\delta p = p -V'(x') \, \delta t  + \delta F(t) 
\end{eqnarray}
As we will see, to spell out the $x$ dependence of the random force, 
we should write  $\delta F(t) = x^{1/3}(t) \eta(t)$, where the 
distribution of the variable $\eta(t)$ is controlled by a function 
$\rho(\eta)$. Now, let us write the distribution function for $x'$ and
$p'$, which by definition is $W(x',p';t+\delta t)$. We get  
\be
W(x',p';t+\delta t)= \int d\eta\:  \int dx\,  dp\ 
\rho(\eta)\,  W(x,p,t)\,  \delta(x'- x- \delta x )\,
 \delta (p'- p -\delta p)
\ee
Now we should eliminate the integrals over $x$ and $p$ with the use of
the delta functions. For that purpose we should make use of the 
time-reversed evolution equations
\begin{eqnarray}
x= x' - \delta t \, \frac{p}{m}\equiv f(x', p', \eta) \:\:\:\:\:\:\:\:\:\:\:\:\:\:\:\:\:\:\:\:\: \\
p= p'+ V'(x') \, \delta t  - x^{1/3}\eta \equiv g(x', p', \eta)
\end{eqnarray}
We get 
\be
W(x', p', t+ \delta t)= \int d\eta\,  \rho(\eta)\ 
W(f(x', p', \eta), g(x', p', \eta), t)\ J(x',p')
\ee
where $J$ is the jacobian of the change of variables. 
Finally, we expand the equation in powers of $\delta t$ and $\eta$. 
Keeping terms linear in $\delta t$ only, we obtain a discretized
version of the quantum Liouville equation provided we demand 
\begin{eqnarray} \label{elang01}
\nonumber
&\int  d\eta\,  \rho(\eta)=1 \:\:\:\:\:\:\:\:\:\:\:\:\:\:\:\:\:\:\:\:\:\:\:\:\:\:\:\:\:\:\:\:\:\:\:\:\:\:\:\:\:\:\:\:\:\:\:\:\:\: \\
&\langle \eta^n \rangle \equiv \int  d\eta\,  \rho(\eta)\ \eta^n=0
\quad \mathrm{ for } \ 0<n\ne 3
\end{eqnarray}
and 
\be
\label{etamom}
\langle \eta^3 \rangle \equiv \int  d\eta\,  \rho(\eta)\ \eta^3=
\frac{\lambda \hbar^2 \, \delta t }{4}
\ee
It is quite obvious that if the  variable $\eta$ is real, the previous
equations are incompatible with a positive definite distribution
function $\rho(\eta)$, since in that case $\langle \eta^{2 n} \rangle\ge 0$
or all moments should be 0. 
Despite this fact in the next section we will see how we can relate
the equation to that of a truly positive-definite distribution
function, for which sampling statistical methods are applicable.

The generalization of the previous formulas to the case of several
variables is fairly straightforward. In principle, the force is now
replaced by a vector of forces $\delta F_i$. These forces are functions 
of several random variables with non-positive definite distribution
functions. A simple way to parametrize these forces is 
\be \label{vecforranf01}
\delta F_i= \eta \chi_i
\ee
where the distribution function of $\rho(\eta)$ coincides with the one
of a single variable. The remaining variables $\chi_i$ 
can be distributed according to  standard positive-definite distributions.

As an example consider the case of the O(N) symmetric potential. In
this case one can write $\chi_i=x_i \tau + \xi_i$ and choose a simple 
positive definite distribution function $\tilde{\rho}(\tau,||\vec{\xi}||)$ 
to recover the time-discretized version of the Wigner-function
equation. We leave the details to the reader.

For the case of a d-dimensional lattice field potential given in the
previous section, a choice like $\chi_i=x^{1/3}_i \xi_i$ will do, provided 
the $\xi_i$ are independent random variables with 
 vanishing average and $\langle \xi^3_i\rangle=1$.

\section{A new computational method}\label{smethod}
In the previous section we have seen how to reproduce the evolution
equation for the Wigner function by means of a Langevin approach with a 
random force with non-positive definite distribution function. This is 
the starting point for a new procedure to approximate the quantum
evolution which we will explain in this section. The method depends on
three steps or approximations which are intimately connected among themselves.
The first part is a coarse-grain approximation in the momenta $p_i$,
with a characteristic coarse-graining parameter $\epsilon$. Next we
will show  how one can reproduce the effect of the non-positive
definite distribution function $\rho(\eta)$ by means of a purely
Markovian process involving ordinary probability measures. This will
allow the use of standard sampling techniques to generate the
distribution. The last step will be the introduction of a parameter 
$\kappa$ multiplying the quantum term in the evolution equation for
the Wigner function. The parameter interpolates between a purely
classical evolution ($\kappa=0$) and the full quantum evolution (for 
$\kappa=1$). The whole procedure can be used to compute the evolution
of quantum expectation values in powers of $\kappa$. This is similar
to an expansion in powers of $\hbar^2$, although part of the $\hbar$
dependence sits in the initial condition and is left unchanged.

As mentioned in the previous paragraph, our  first step  is a coarse-grain
approximation, which will amount to an approximation to the 
non-positive definite $\rho(\eta)$. For that purpose, we   
might relax the condition that $\langle \eta^n \rangle=0$ for
$n>3$. One possible realization of the conditions is achieved 
by the following family of distribution functions:
\be
\rho_N(\eta) = \delta(\eta)+ \sum_{i=1}^{M_N} \frac{\gamma_i}{\epsilon^3}
(\delta(\eta-\epsilon \alpha_i) -\delta(\eta+\epsilon \alpha_i))
\ee
where $\gamma_i$ are positive numbers, $\alpha_i$ are real values, and
$\epsilon$ is a free parameter. Although not explicitly indicated, the
coefficients $\gamma_i$ and $\alpha_i$ do depend on N. They are
determined by imposing that all even moments
vanish and odd moments, given by
\be
\mu_{2 p +1} \equiv \langle \eta^{2 p + 1} \rangle_{N}= 2 \sum_{i=1}^{M_N} 
\gamma_i \, \alpha_i^{2p+1} \epsilon^{2(p-1)}
\ee
vanish for $p\le N$, with the exception of $p=1$ given by
Eq.~\ref{etamom}. If we take, without loss of generality, that the
parameters $\alpha_i$ are of order 1, this condition implies that 
\be
\gamma_i  \propto \frac{\lambda \, \hbar^2 \, \delta t }{8}
\ee
In general, the solution to the set of constraints will not determine 
the parameters $\gamma_i$ and $\alpha_i$ uniquely.
This freedom  in the choice of the parameters  is a bonus, since one can 
test the effects of the coarse-graining
on the results, by exploring   different choices. Furthermore, a
better approximation is obtained by taking larger values of $N$, which
will be referred as the level of the approximation. The number of terms 
$M_N$ has to grow as $N$ grows. An alternative method to improve the
accuracy would be to reduce  the value of $\epsilon$, thus reducing the
effect of higher order derivatives of  the Wigner function. As we will
see later, there is a limitation to the minimal value of  $\epsilon$, 
which is dictated by the range of time for which the method 
would be applicable.

The next ingredient will be that of relating  the Wigner function to 
a positive-definite  distribution function, which can be
approximated by samples. The same can be done for the distribution function 
$\rho(\eta)$ as follows. Let us introduce a positive-definite function 
$\hat\rho(\eta,\sigma)$ involving a discrete Ising-like variable $\sigma=\pm 1$.
This function is related to $\rho(\eta)$ by 
\be
\rho(\eta)= \mathcal{N} \sum_{\sigma = \pm 1} \hat\rho(\eta,\sigma) \, \sigma 
\ee
The function   $\hat\rho(\eta,\sigma)$ is normalized
as a probability distribution
\be
\sum_\sigma \int d\eta \, \hat\rho(\eta,\sigma) =1
\ee
and hence the prefactor is given by:
\be
\frac{1}{\cal N} = <\sigma>_{\hat\rho}\, \equiv  \sum_\sigma \int d\eta \,  
\hat\rho(\eta,\sigma) \, \sigma
\ee

The construction can be extended  to the coarse-grained version given before. 
Hence, we define 
\be
\hat\rho_N(\eta,\sigma)=\frac{1}{\cal N} \left[ \frac{1+\sigma}{2} \, \delta(\eta)+
\sum_{i=1}^{M_N} \frac{\gamma_i}{\epsilon^3} \, \delta(\eta-\epsilon
\alpha_i\sigma )\right]
\ee
The normalization condition implies
\be
{\cal N}= 1+2\sum_{i=1}^{M_N} \frac{\gamma_i}{\epsilon^3} 
\ee

A similar procedure can be applied to the Wigner function 
\be
W(x,p;t)= K(t) \sum_{\sigma=\pm 1} \hat W(x,p,\sigma;t) \, \sigma 
\ee
where $\hat W(x,p,\sigma;t)$ is a well-defined probability distribution. 
This is equivalent to writing the Wigner function as the difference of
two positive definite functions. If we label by $\langle \
\rangle_{\hat W}$
the expectation values with respect to $\hat W(x,p,\sigma;t)$, then the 
expectation values with respect to the Wigner function are given by 
\be
\label{expect}
\int d x\, d p \ O(x,p) \, W(x,p;t)= \frac{\langle \sigma \, O(x,p)
\rangle_{\hat W}}{\langle \sigma   \rangle_{\hat W}}
\ee
Now, one can obtain a time-discretized 
evolution equation for $\hat W(x,p,\sigma;t)$ involving $\hat\rho(\eta,\sigma)$
as follows: 
\be
\label{new_evol}
\hat W(x,p,\sigma;t+\delta t) = \sum_{\mu=\pm 1} \int d\eta \,
\hat\rho(\eta,\mu)  \hat W(x-\delta x ,p-\delta p, \mu \cdot \sigma;t)
\ee
where the displacements $\delta x$-$\delta p$ are those coming from
the classical equations of motion with a force proportional $\eta$ (as
before). Summing the previous equation over $\sigma$ and using the 
previous definitions, we re-obtain the discretized evolution equation for
the Wigner function provided 
\be 
K(t+\delta t)= {\cal N} K(t)
\ee
We see that this implies that the normalization factor grows
exponentially, and this is precisely the main numerical limitation of
the method. 

Having set up an evolution equation for the probability distribution 
$\hat W(x,p,\sigma;t)$ we can employ standard sampling techniques to
approximate it. This leads us to the concept of  {\em signed} samples:
a collection of ${\cal M}$ points $\{x^{(a)}, p^{(a)}, \sigma^{(a)}\}$ such that 
\be
\int d x \, dp\, \sum_{\sigma=\pm 1} \hat W(x,p,\sigma;t) \, O(x,p;\sigma)
\approx \frac{1}{\cal M} \sum_{a=1}^{\cal M} O(x^{(a)},p^{(a)};\sigma^{(a)}) 
\ee
From Eq.~\ref{new_evol} one can determine the time evolution of samples 
given a realization of the noise $\eta$. This is a Markovian process
where $\hat\rho(\eta,\mu)$ determines the conditional probability for a
transition from one point in phase-space to a new one and a possible
change of sign of the discrete Ising variable. 

If we use the coarse-grained distribution $\hat\rho_N(\eta,\mu)$,
then with a certain probability we would simply evolve the system with 
the classical equations of motion, and with probability proportional to
$\delta t$ we  would produce a jump in the value of momentum and a possible 
flip of the sign of $\sigma$. The reason for referring to the first
approximation as coarse-graining in momentum space, has to do with the
discrete magnitude of the jump (of order $\epsilon$) at each step of
the time-evolution. If the Wigner function would be a polynomial in
$p$, the approximation would be exact. 

If we start the evolution with a positive definite Wigner function,
then $\frac{1+\sigma}{2}W(x,p,t=0)= \hat W(x,p,\sigma,t=0)$. Hence,   
all points in the sample have $\sigma^{(a)}=1$. As time evolves 
some of the points acquire a negative value of $\sigma^{(a)}$. At time 
$t=n\delta t$ the probability that a point in the sample has negative 
$\sigma$ is given by 
\be
P_-= \frac{1-  \langle \sigma   \rangle_{\hat W}}{2}= 
\frac{1}{2}\, (1-(<\sigma>_{\hat\rho})^n) \approx \frac{1}{2}\, (1- e^{-A t
\lambda \hbar^2/\epsilon^3} )
\ee
with $A$ a number of order 1, which depends on the detailed form of
$\hat\rho_N$. The probability  approaches $1/2$ exponentially in time. Once  this
happens we encounter a severe sign problem in computing averages
according to the formula Eq.~\ref{expect}, since the  averages
result from strong cancellations from $\sigma=1$ and $\sigma=-1$. At
this point the method breaks down. The typical time when this happens
is given by 
\be 
T \approx \frac{\epsilon^3}{\lambda \hbar^2}
\ee
Clearly things get worse as we decrease $\epsilon$ and improve as we 
decrease $\lambda$ and $\hbar$. However, the  magnitude of $\epsilon$
must be related to the  typical range of variation in $p$ of the Wigner 
function. Otherwise, corrections involving higher order derivatives of
the Wigner function become large. The systematic  errors associated 
to the coarse-graining can be checked by varying $\epsilon$ or by
using different choices of $N$. For small enough  $\epsilon$, 
the approximation should behave better for larger $N$. 

Once the maximum acceptable value of $\epsilon$ is selected, 
the method stated before only allows the computation of 
quantum expectation values for a range of times. This limitation,
although important, does not destroy the usefulness of the method.
Time-range  limitations are already present in studying the classical evolution
equation of a quantum field theory discretized on a lattice. The moment 
that the fluctuations become sizable at the  cut-off scale, the
discretized evolution equation fail to reproduce the continuum 
evolution. Fortunately, in many applications a good part of the interesting
physical processes take place in a relatively short period of time. 
This is the case, for example,  in the context of preheating in the
early universe, which is one  of the main motivations that we had for
embarking in the present work. 

In situations in which a reliable full quantum evolution can only be
carried for a too short lapse of time, the method can be used to 
compute first-order quantum corrections as follows. This is the third
ingredient that we anticipated in the first paragraph of this section. 
The strategy is to consider a modified evolution equation with a new
parameter $\kappa$ multiplying the last term of Eq.~\ref{main_eq}. 
This can be interpreted as multiplying $\hbar$ by $\sqrt{\kappa}$. 
In this fashion one extends the time range of applicability of our
method by the  multiplicative factor  $1/\kappa$. Now, evolving the
system for several other smaller values of $\kappa$, one can determine 
the evolution of the expectation values as a power series in $\kappa$. 

A practical problem concerns accuracy. Since, our goal is to estimate 
the quantum fluctuations, we realize that when reducing $\kappa$, they
have  decreased by the same factor. To maintain the signal to noise ratio
one should increase the size  of the sample by $\kappa^2$, which would 
increase the computation time by the same factor. Since the reduction
in $\kappa$ was dictated by the necessity to extend the range in time 
of the simulation, we conclude that this can be done at a cost in
computer time which only grows polynomially with this range.

Of course, the drawback is that one does not compute the full quantum 
effects but only to leading order in $\kappa$ (i.e. $\hbar^2$). 
This by itself is an important result because it would
serve as a measure of the size of quantum effects and as a next-to-leading
correction to the classical approximation. Higher order powers of 
$\kappa$ are also computable at a  higher cost in computer time.

To test these ideas we decided to study  several simple quantum mechanical
systems for which the exact quantum evolution can be determined by
numerical integration of the Schroedinger equation. The results will
be presented  in the next section and compared to other extensions of
the classical approximation given in the literature. 

Before closing this section, we should comment on the modifications 
necessary to extend the procedure defined previously to the case of
many degrees of freedom. Our method might not be optimal for the case
of very few degrees of freedom, however, it was designed to be
extensible in an affordable way to the case of many degrees of freedom. 
In this respect it differs from other proposals in the literature 
based on histogramming which seem impossible to extend to a large
number of variables. 

The formal extension of all the approximations involved in the
method to the case of several degrees of freedom is indeed trivial. 
In the way that the Langevin process was extended at the end of the
last section, it turns out that there is a unique {\em random variable}
$\eta$ having a non positive definite distribution function with
similar or exact properties as the one appearing for the one-variable
case. Thus, the coarse graining $\rho\longrightarrow \rho_N$ and 
extension to positivity $\rho(\eta)\longrightarrow \hat\rho(\eta,\sigma)$
remains exactly the same irrespective of the number of variables. 
The rest of random variables entering the force are of the
conventional type and their number increases linearly  with the number
of degrees of freedom, and so does the computation time for a given
time-step. The sample now, however,  involves   trajectories in the
multidimensional state of the system. This is exactly the same as for 
the classical evolution (with random initial conditions), except that 
now there is a single  additional Ising-like variable $\sigma$. Thus,
for a fixed number of trajectories the computational cost will grow
in a similar fashion to that of the classical evolution. 
A new question to worry  about is whether the number of trajectories 
needed to obtain results with a reasonable accuracy depend upon the 
number of degrees of freedom. This will be studied in the next chapter 
for one particular example.

\section{Testing the classical approximation and its extensions}
In this section we will present the results of our tests 
of the classical approximation and of  our proposed method to 
obtain quantum corrections, together  with other proposals. The first cases 
are particularly simple situations with one or two degrees of freedom
for which the exact  quantum evolution can be obtained through 
the numerical integration of the  Schroedinger equation. Later 
we will explore the first steps towards a possible application to 
quantum field theory. 

Our first example  will be a simple anharmonic oscillator at intermediate
values of the self-coupling. The potential is that given in 
Eq.~\ref{potential} with the following choice of parameters:
\be \label{anarosparam}
m= 1 \quad ; \quad \quad \mu^2= 0.5 \quad ;\quad \quad \lambda= 0.45 
\ee
The value of the dimensionless parameters  mentioned in Section 2 are
given  by $r=s=1.57$. We choose as initial condition a gaussian pure state with width given 
by $\sigma \mu^{2/3}=0.45/2^{1/3}$. Our main observable 
was taken to be the expectation value of the square of the position
operator  $Q^2$ as a function of time, which is noted $\langle Q^2\rangle(t)$.
The Hamiltonian, the initial state and the observable were used in a 
previous paper with a similar spirit to ours~\cite{Aarts:2006cv}. However, for
illustrative purposes  we are presenting the results for  a higher value of  the
self-coupling $\lambda$, for which quantum effects are stronger.

\begin{figure}
\begin{center}
 \subfigure[]{
\includegraphics[angle=-90, width=0.48\textwidth]{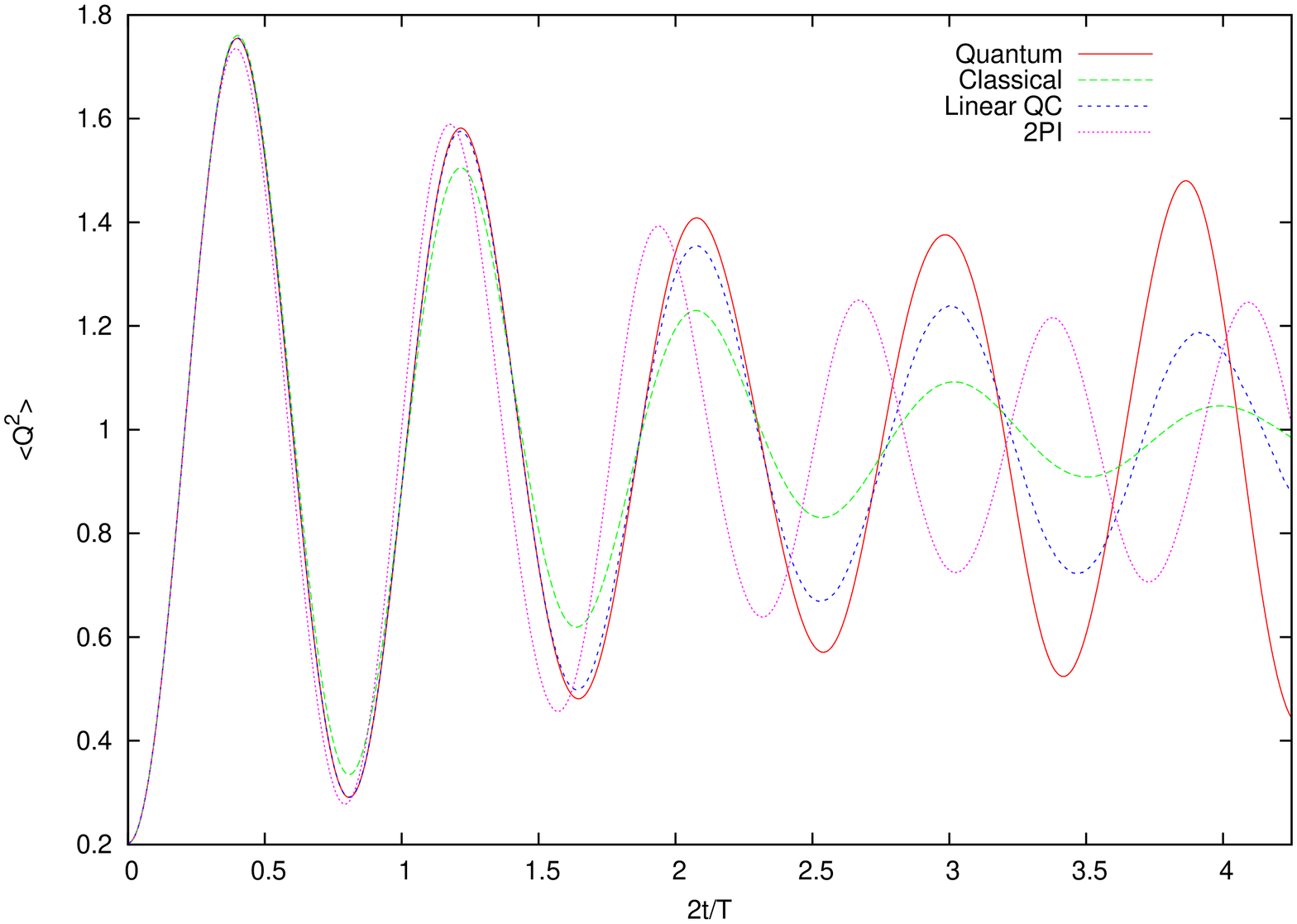}
\label{fig2a} }
 \subfigure[]{
\includegraphics[angle=-90, width=0.48\textwidth]{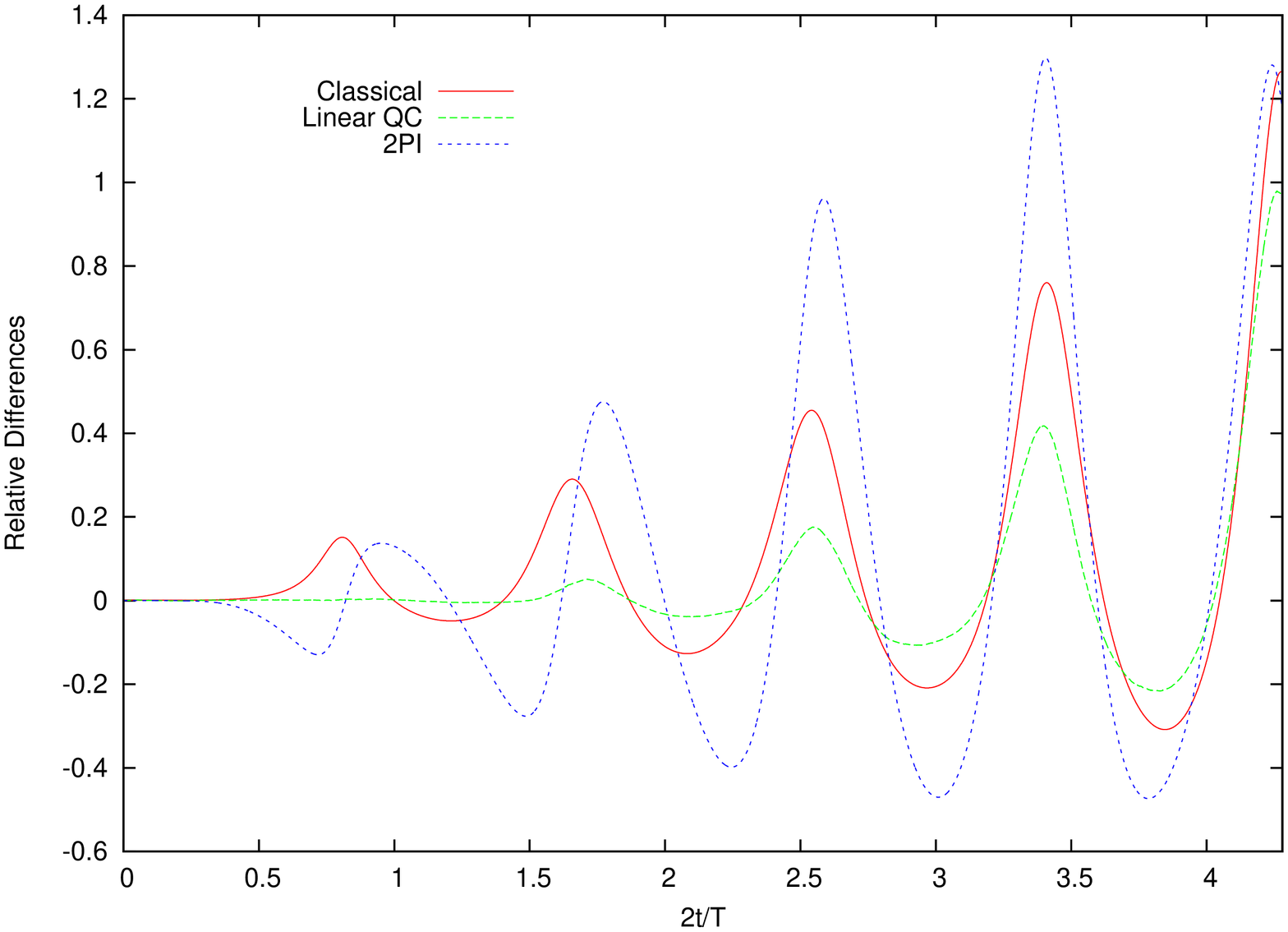}
\label{fig2b} }
\caption{(a): Time evolution of $\langle Q^{2}  \rangle$ compared to
the classical approximation, the 2PI truncation to NLO order in 1/N
expansion, and the method
proposed in this paper.  (b): Relative error committed in each of the
approximations as a function of time.}
\label{fig2}
\end{center}
\end{figure}

The main results are collected in Fig.~\ref{fig2a}-\ref{fig2b}.
In the first figure the time evolution of the observable is displayed in 
units of the half-period of the $\lambda=0$ system. 
This expectation value performs oscillations with a frequency close 
to that of the $\lambda=0$ system. The classical approximation, also
displayed in the figure, oscillates as well, but the amplitude gets  damped 
very fast with time. This damping is a typical feature of the classical
approximation which has been pointed out repeatedly (including 
Ref.~\cite{Aarts:2006cv}).
The figure also shows two other curves. The first being the 2PI
approximation obtained by  keeping only the leading and next-to-leading (NLO) 
diagrams in a 1/N expansion~\cite{Berges:2001fi}. Notice that in 
this case  the amplitude of the oscillation is not decreasing, but there is
a shift in the period oscillation. This might not be a
serious drawback in extracting average properties over time.
Finally, we also present the result of the new method explained before,
which includes  the classical approximation and the leading $\hbar^2$
correction. The exact details are explained below.

In Fig.~\ref{fig2b} we present  relative error of each approximation, 
namely the difference between the quantum evolution and the corresponding
approximation, divided by the quantum result. The first line
corresponds to  the classical approximation, which oscillates with increasing
amplitude. Quantum corrections start being negligible and grow 
to a 20\% level at $2t/T \approx 0.8$, and  40\% level at $2t/T \approx 2.5$. 
The 2PI curve appears  to be the  worst, but this is due
to the shift in period with respect to the quantum curve. A more fair
presentation  should involve a comparison of the height of the maxima,
for which 2PI is certainly better than the classical approximation.
The last curve is our  calculation including quantum effects up to linear 
order in $\hbar^2$, using the method described in the previous section. 
Notice, that it provides   a very accurate description  of the quantum effects
up to $2t/T\approx 1.4$. Beyond this point  it has more sizable errors, but
certainly smaller than the classical approximation. Furthermore, it
provides at least  an estimate of the errors committed by employing the classical
approximation. 

The actual procedure that we followed to determine the leading 
quantum correction (LQC) is the following. We re-scaled the size of the
quantum term of quantum Liouville equation by using $\kappa$.  
Then,  we  used the coarse-grained  approximation to $\rho(\eta)$ up to the 
second level ($N=2$,  $\langle
\eta^5\rangle=0$ but $\langle \eta^7\rangle \ne 0$), and the sampling
method described in the previous chapter, to study the quantum evolution 
equation for a given value of $\kappa$.
The formula to obtain the approximation (LQC) including the 
contribution linear in $\hbar^2$ to any observable $O$ is 
\be
O_{\mathrm{LQC}}= O_{\mathrm{clas}}+\frac{1}{\kappa}(O_{\kappa}-
O_{\mathrm{clas}})
\ee
The curves depicted in Fig.~\ref{fig2} were obtained 
using $\kappa=1/6$. 

\begin{figure}
\begin{center}
\includegraphics[angle=-90, width=0.7\textwidth]{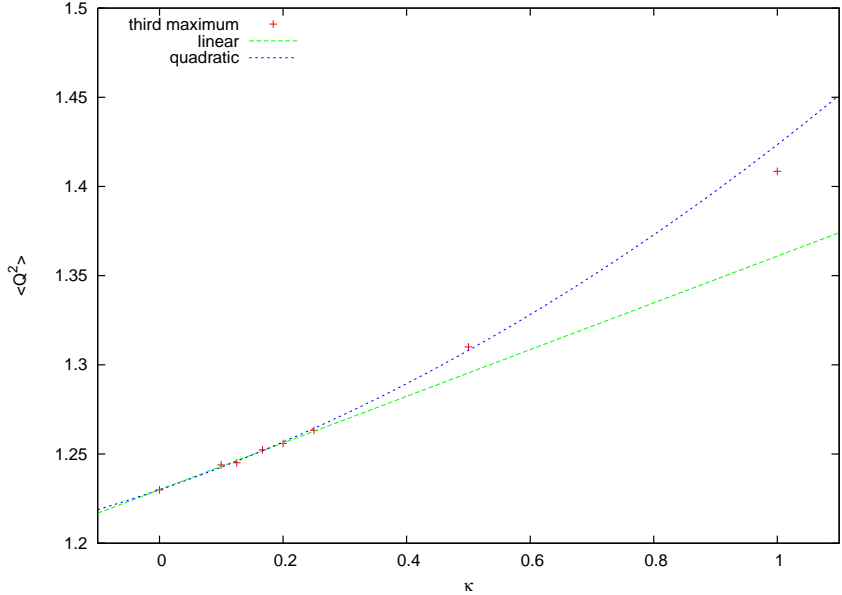}
\caption{The value or  $\langle Q^{2}  \rangle$ at the third maximum
in time computed using several values of $\kappa$ (see text). The lines
are linear and quadratic fits.}
\label{fig3}
\end{center}
\end{figure}

To check whether $\kappa=1/6$ is  in the linear regime, and to 
give an estimate  of the size of the higher order terms 
in $\hbar^2$, we repeated the procedure and obtained 
$O_{\kappa}$ for several  values of $\kappa$ 
($\kappa = 0, 1/10, 1/8, 1/6, 1/5, 1/4, 1/2$).
  In this way we get an idea about how the value of the observable 
interpolates between the classical ($\kappa=0$) and the quantum value
($\kappa=1$).  In  Fig~\ref{fig3} we display the result obtained 
for the expectation value of $Q^2$ at the position of the third maximum
($2t/T=2.1$). The y-axis gives the values obtained for the different
values of $\kappa$ mentioned previously. We also display the 
value for the classical approximation ($\kappa=0$) and the
full quantum result ($\kappa=1$). It is quite clear that the results follow an approximate 
linear dependence for $\kappa<1/3$. The straight
line is the result of a linear fit (1 free parameter) in
this range. The extrapolation to $\kappa=1$ of the straight line is very approximately our
estimate of the quantum evolution up to next-to-leading order in
$\hbar^2$ (the leading order being the classical approximation). 
In this particular case we see  that the linear quantum correction
term  (LQC) accounts for  80\% of the quantum effects. 
Thus, with the addition of the classical approximation, we reproduce
the exact quantum result with a 3\% error. 
In principle, one  could go beyond the linear approximation and determine higher order 
corrections in $\hbar^2$. If we add  the result of $\kappa=0.5$ to the 
data and fit the results to a second degree polynomial in $\hbar^2$ we 
get an even better approximation to the quantum result (second line 
in Fig.~\ref{fig3}). 

Although the previous results by themselves show that our procedure
cannot be completely misguided, we have analyzed the different sources of error in the
determination of the $\hbar^2$ correction presented above. Using
jack-knife methods we can quantify the purely statistical errors.
They increase with time but remain always at the level of a few
percent. A much more difficult estimate is the effect of the
coarse-graining in momenta. This can be estimated by changing the value 
of $\epsilon$ and/or adding more terms in the discretization to impose 
$\langle \eta^{2M+1}\rangle=0$. In particular, we have used results 
at $\epsilon=0.3$ and three levels of discretization. The effect is
more pronounced at the maxima and minima of the oscillations and the
better  the approximation, the  closer the results to the actual 
quantum evolution. We estimate that, at most,  errors (to the quantum correction)
could be of the order of 10-15\% at the third maximum ($2t/T=2.1$) rising up
to 20-25\% at the fourth maximum ($2t/T=3.0$). The same conclusion
follows both by comparison of the different levels as well as by
extrapolation in $\epsilon^\alpha$ (with $\alpha\approx 3$) to $\epsilon=0$. 
The conclusion
is that, even if the errors are sizable, the method provides a good
estimate of the quantum effects. 

\begin{figure}
\begin{center}
\subfigure[]{
\includegraphics[angle=-90, width=0.48\textwidth]{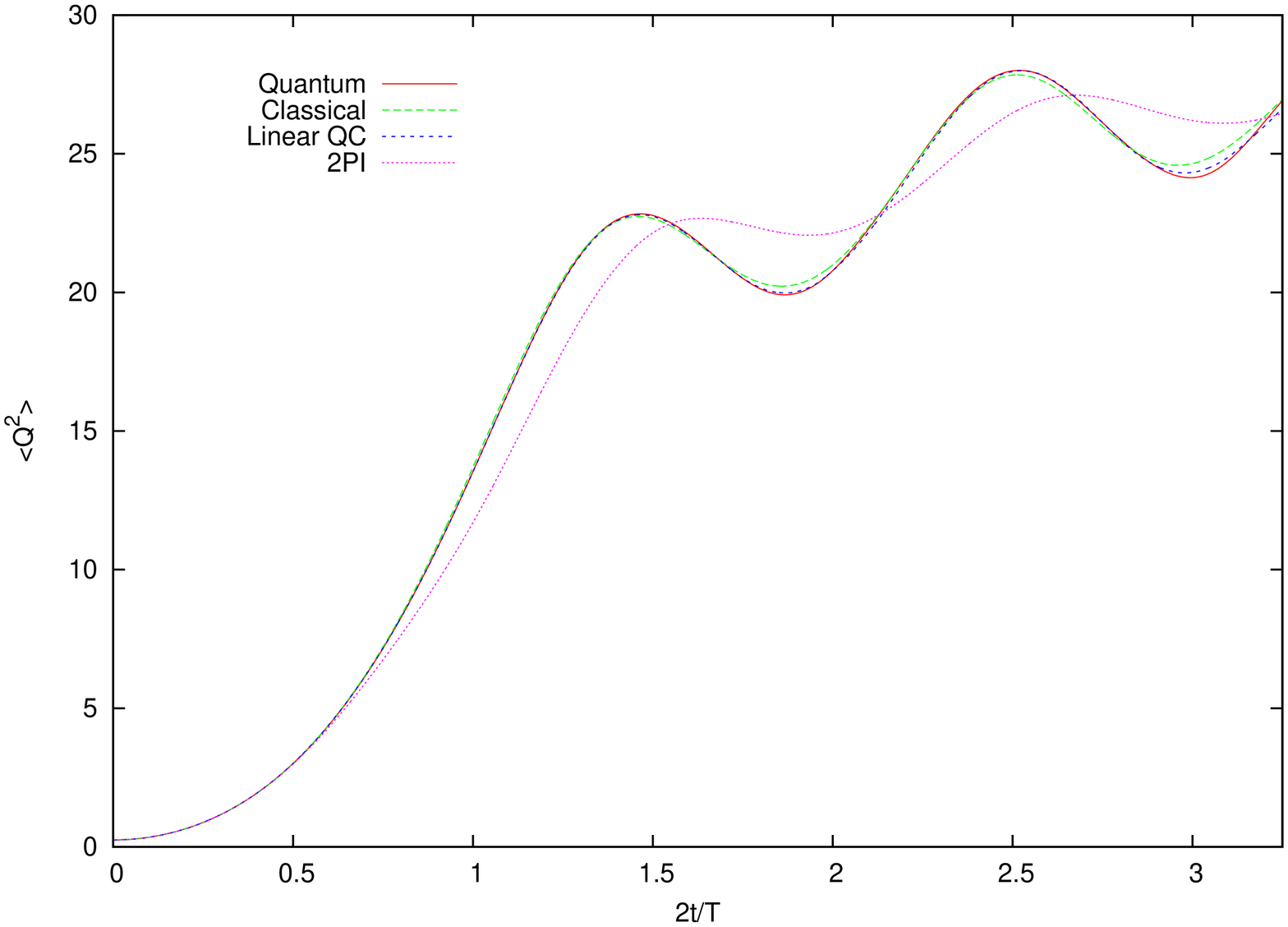}
\label{figth01} }
\subfigure[]{
\includegraphics[angle=-90, width=0.48\textwidth]{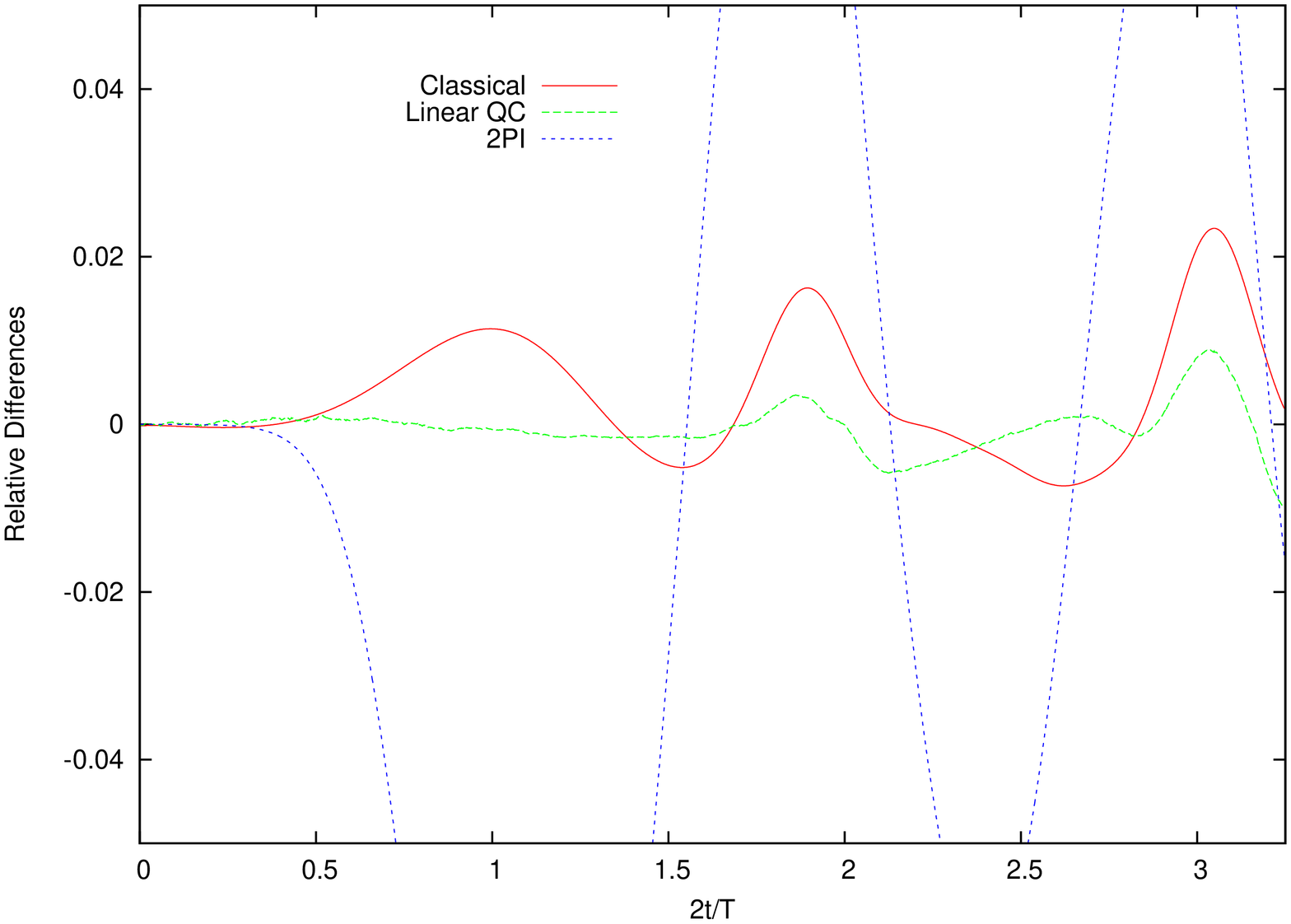}
\label{figth02} }
\caption{(a): The same as Fig.~\ref{fig2a} but for the potential in
Eq.~\ref{tanhpotential}. (b): Relative differences as in
Fig.~\ref{fig2b}.}
\label{figth}
\end{center}
\end{figure}

Before embarking into the generalization to several variables, we
tested the situation for another case having several distinct features
which are present in some of the phenomenological applications to
cosmology. We considered a potential of the form 
\be
\label{tanhpotential}
V(x,t) = - \frac{1}{2} \mu^{2} \tanh(\alpha t) x^{2} + \frac{\lambda}{4!} x^{4}
\ee
with the parameters chosen to be 
\be
\mu^{2} = 2 \quad ; \quad \quad \alpha = 0.2 \quad ; \quad \quad
\lambda = 0.4
\ee
This time-dependent potential  provides a smooth interpolation between a single-well and a
double-well potential mimicking the situation occurring in hybrid inflationary
models.  Notice that  tunneling effects are possible now. 
The initial condition is a pure state given by a gaussian with $\sigma \mu^{2/3}
= 1/4^{1/3}$.  The same observable as before, $\langle Q^{2}  \rangle$, is displayed
in Fig.~\ref{figth01} for a range of times for which the potential has
basically evolved to the  future asymptotic potential. Hence, as expected,
the   expectation value
migrates from its  initial value to performing oscillations around 
$x_{min}^{2}$, where $x_{min}$ is the minimum of future asymptotic potential.
Also shown, is the corresponding curve for the same  2PI approximation
mentioned earlier. Although following the pattern of the quantum
result, the differences are substantial.   On the other, hand the classical evolution 
works  quite well for this case. However, the addition of the
Linear Quantum correction (LQC) using our method (with $\kappa=1/2$) makes the result much
better, as can be  seen  when looking at Fig.~\ref{figth02}, where we display the relative
differences with respect to the quantum evolution as before.  

A full comparison of competitive methods suggested us to include the
results of the Complex  Langevin method described
in~\cite{Berges:2005yt}, and we invested considerable effort in doing
so. The method generates a collection of histories as a function of an
additional time variable. Our naive implementation, however,  led to a 
growing number of trajectories blowing up in this additional time. It
is easy to see that this is a feature of the complex evolution in the 
discretized new variable and in the absence of random noise. Both the 
problem and the possible cures have been documented in the literature
of the subject. One can employ more refined discretizations or use 
a much smaller Langevin step~\cite{Berges:2005yt}, but this pays 
an obvious price in  computational cost. A way out proposed in
Ref.~\cite{Aarts:2009dg} is to use an adaptive stepsize. Another
possibility is a modification of the noise~\cite{Aarts:2009uq}. Using 
these techniques we were able to eliminate most of the divergent
trajectories. A small fraction, but growing in extra-time, remained. 
We took the attitude mentioned in Ref.~\cite{Berges:2005yt} to discard 
them or go back in time and re-evolve them. 

Another problem mentioned in the literature is the issue of convergence.
Sometimes the system can show  a limited decay to the equilibrium distribution 
 or can converge to a wrong limit~\cite{Aarts:2009uq}
\cite{Aarts:2011ax}. This seems to be case in our tests of the
previous examples. Furthermore, the results obtained seemed robust
under changes of the methodology (adaptive step, different stepsizes,
discarding or re-evolving) and stable under further evolution in the
additional time. These results did not even show the right oscillatory
pattern already seen in the classical approximation. 
Some  authors~\cite{Aarts:2011ax}  claim that to ensure the correct
convergence other modifications are  necessary. However,  given that
this was not the main issue of this paper, we decided to drop out these
results and defer its study to further scrutiny.
 
\subsection*{Extension to several variables}
Since our ultimate goal is that of studying the quantum evolution of
fields, it is crucial to determine how does the new method that we have 
presented depend upon the number of degrees of freedom. The standard 
non-perturbative treatment of quantum fields proceeds through a
lattice discretization and a subsequent continuum limit.
Renormalization is a crucial ingredient in the process to obtain
meaningful physical results. The latter aspect lies somewhat far 
from the scope of this paper and will be addressed in a future
publication. The focus here is rather upon the numerical feasibility of
the procedure. For attaining  acceptable results one has to reach
a number of variables within the range of those customary for these 
kind of simulations. A priori, the method presented here is 
capable of doing so, since its computational effort is comparable 
to that involved in the classical approximation for a given number of
sampling trajectories, which has  been used successfully in this context. 

However, there is a point of concern which we want to address. It
might well happen that the number of trajectories needed to attain 
a given precision in the estimation of the quantum corrections grows
with the number of degrees of freedom: A polynomial growth is
acceptable, an exponential growth is not.

As
a testing example we have considered a lattice version of a two-dimensional 
scalar field theory which has been studied by other authors in this
same context~\cite{Salle:2001xv}. We take a real scalar quantum field in $1+1$ dimensions with Hamiltonian
\begin{equation} \label{epqf01}
H(t) = \int dx \Big[ \frac{1}{2} \pi^{2}(x,t) + \frac{1}{2} (\nabla \phi(x,t))^{2} + \frac{1}{2} \mu^{2} \phi^{2}(x,t) + \frac{\lambda}{24} \phi^{4}(x,t) \Big]
\end{equation}
where $\mu^{2}$ may depend on t. We have already explained the
influence of $\hbar$ factors, so from now on we assume 
natural units, $\hbar=c= 1$. 
The dimensionless field $\phi$ and its conjugate momentum $\pi = \dot{\phi}$ 
satisfy equal-time canonical commutation relations
\begin{equation}
[\pi(x), \phi(y)]= -i \, \delta(x-y)
\end{equation}
which gives $\pi(x)$ dimensions of inverse length.

The next step is to consider the lattice version of the previous
Hamiltonian.  Continuous space is approximated by a discrete number of
points $x_{n}=na$  separated by a distance $a$, the lattice spacing. To deal with a 
finite number of variables we must, in addition, put the system in a
box of size $L$ with periodic boundary conditions. Altogether, we end
up having $N=L/a$ variables $\phi_{n}(t) \equiv \phi(na,t)$. The
corresponding conjugate momenta $\pi_{n}(t)$ satisfy the commutation
relations
\begin{equation}
[\pi_{n}, \phi_{m}]= -\frac{i}{a} \, \delta_{n m} 
\end{equation}
where the factor of $a$ is necessary to preserve the dimensions of the
conjugate momentum.  Naive discretization then leads to the Hamiltonian 
\begin{equation} \label{epqf03}
H = \sum_{n=0}^{N-1}a \Big[ \frac{1}{2} \pi_{n}^{2} + \frac{1}{2} (\nabla \phi_{n})^{2} + \frac{1}{2} \mu^{2} \phi_{n}^{2} + \frac{\lambda}{24} \phi_{n}^{4}  \Big]
\end{equation}
with $\nabla \phi_n=(\phi_{n+1}-\phi_{n})/a$. 
After a suitable rescaling of the variables  and of the
parameters the Hamiltonian can be cast in the form Eq.~\ref{ewigfields01}.

After presenting our system and its discretized version, let us
consider the dynamical process that we will study following
Ref.~\cite{Salle:2001xv}. The idea is to study the evolution of the
system after a {\em quench}. In practice, this means that the $\mu^2$ 
parameter flips its sign abruptly at time $t=0$ 
from  a positive value to a negative one. This can be seen 
as a limiting version  of our previously smooth ($\tanh$) transition 
from single to double-well.

In practice, what we will consider is the evolution of the system
for positive times starting (at $t=0$) from an initial state given by
the ground state of the Hamiltonian with $\lambda=0$ and positive
$\mu^2$. This initial state is therefore gaussian as in previous
examples, and is easily generated. For our numerical simulation we
have taken the parameters of the model to be 
\begin{equation}
\lambda=3 m^2 \quad \quad a=0.8/m
\end{equation}
where the unit of mass $m$ is given by $m=\sqrt{-2 \mu^2}$. 
Then we have studied this model for  $N=2$,$4$,$8$,$16$ and $32$.

For a real quantum field theory application one has to study the limit 
of $a$ going to $0$, with a suitable tuning of the parameters dictated
by the renormalization conditions. For very small lattice spacings 
one might even encounter problems of critical slowing down, but these 
will affect other methods too. Here, we will be content with scaling
the number of degrees of freedom  and focusing on statistical
significance and computational load alone. 

Finally, let us select our observables and present our results. Rather
than working with the field variables we can work with their Fourier
modes $\hat\phi_{k}(t)$, since they decouple in the $\lambda=0$ limit.  
We use the following normalization for the discrete Fourier transform
\begin{equation} \label{epqf04}
\hat\phi_{k}(t) = \frac{\sqrt{L}}{N} \, \sum_{n=0}^{N-1} e^{-i \frac{2
\pi}{N}nj} \, \phi_{n}
\end{equation}
where $k=(2 \pi /L)j$ for $j=-N/2+1,-N/2+2,{\ldots} .,N/2$. A similar
expression applies for the modes $\hat\pi_{k}(t)$. Reality of our original
field implies $\hat\phi_{k}^{*}(t)=\hat\phi_{-k}(t)$ (and the same for
$\pi$). 

As mentioned previously for $\lambda=0$ and positive $\mu^2$ all the
modes oscillate with a characteristic frequency $\omega(k)=
\sqrt{4\sin^{2}(j \pi /N)/a^{2} + \mu^{2}}$. 
At the classical level, the flip in sign of $\mu^2$ produces that the 
low lying modes  acquire an imaginary $\omega(k)$, and start growing 
exponentially. In the presence of a non-zero $\lambda$ the  growth ceases
once the non-linear effects become important.

\begin{figure}
\begin{center}
\includegraphics[angle=-90, width=0.7\textwidth]{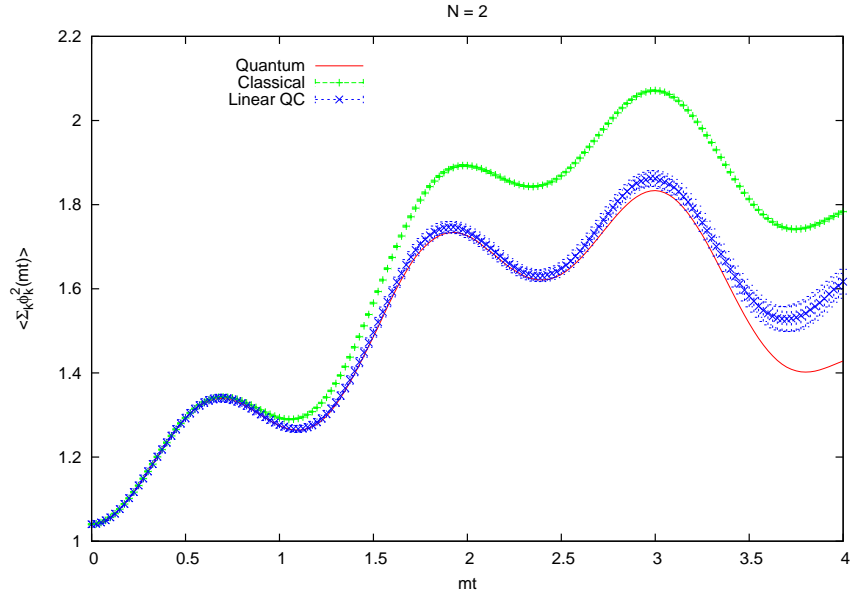}
\caption{Time evolution of $\langle \sum_{k} |\phi_{k}|^{2}(mt) 
\rangle$ for $N=2$, for the fully quantum, classical approximation and 
LQC method for $\kappa = 1/3$.}
\label{figN02b01}
\end{center}
\end{figure}

In Fig.~\ref{figN02b01} we present our results for the sum of 
expectation values of the square of each mode $\langle \sum_{k} |\hat\phi_{k}|^{2}(mt)  \rangle$ for two degrees of freedom $N=2$.
As a matter of fact this observable is just a discretized version of $\int dx \, \phi^2(x)$.
The time evolution as a function of $mt$ is given, as obtained from the
numerical integration of the $N=2$ Schroedinger equation. The result
of the classical approximation and of our LQC method  obtained from $\kappa = 1/3$
are also shown. The exact numerical integration has  negligible errors 
at the scale of this and the following figures. The errors of the remaining
approximations were obtained by applying a jack-knife method to the sample of
trajectories. The total number of trajectories used for this data was 
$\mathcal{M}=8\times 10^{6}$.

\begin{figure}
\begin{center}
\subfigure[]{
\includegraphics[angle=-90, width=0.48\textwidth]{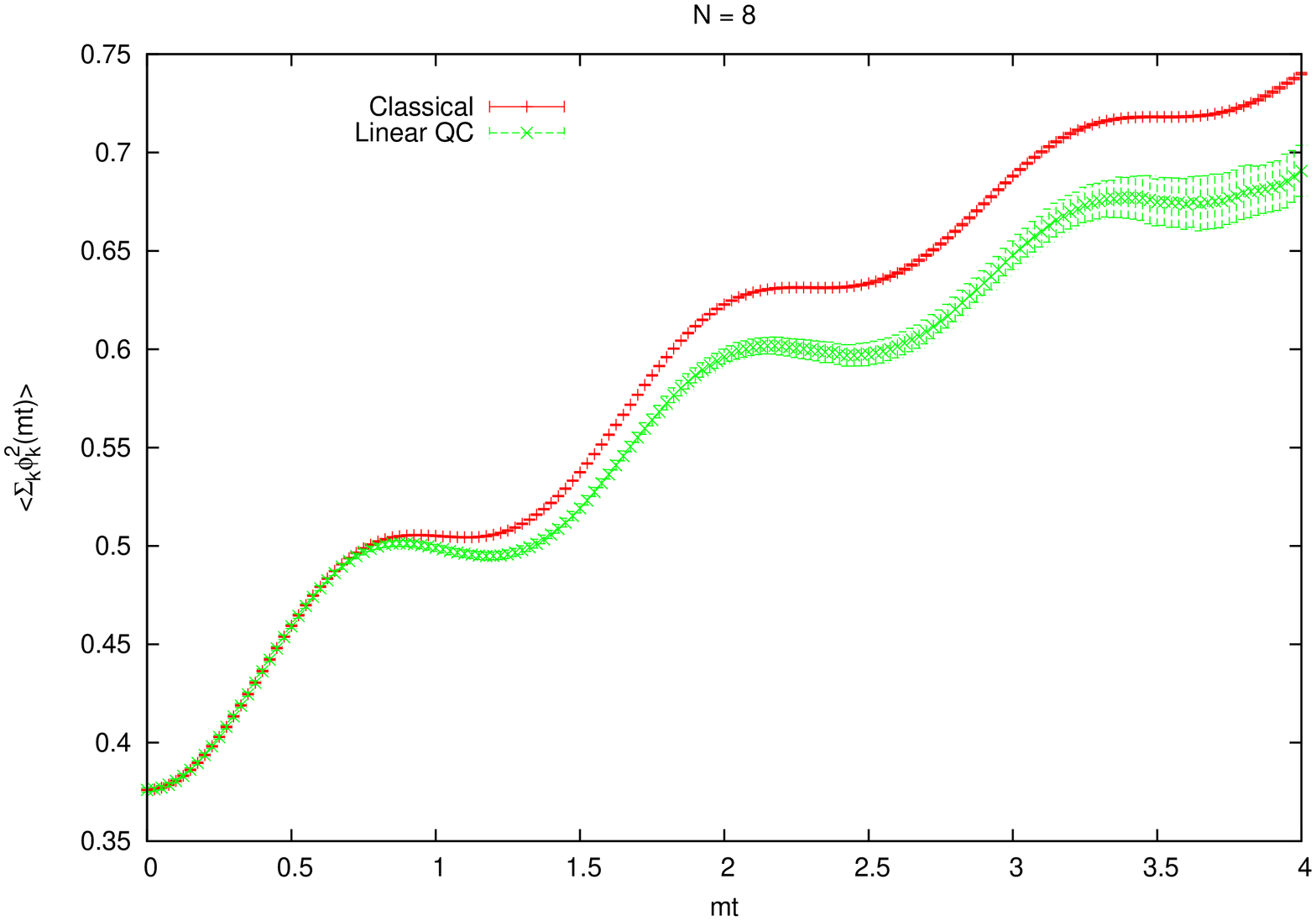}
\label{figN08b01} }
\subfigure[]{
\includegraphics[angle=-90, width=0.48\textwidth]{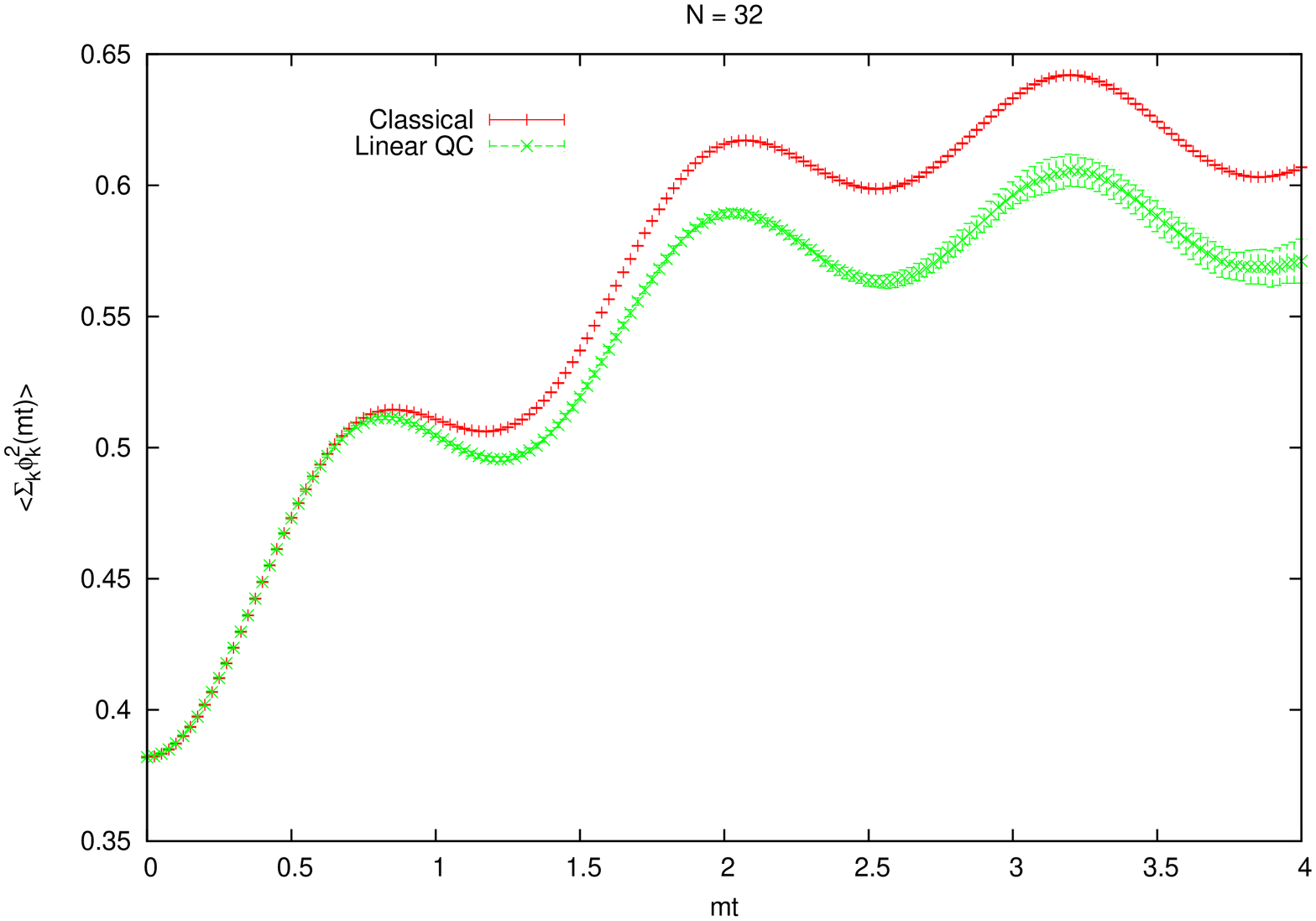}
\label{figN32b01} }
\caption{The same as Fig.~\ref{figN02b01} but for $N=8$ (a) 
and  $N=32$ (b). }
\label{figN08N32}
\end{center}
\end{figure}

The results show the same pattern as before. The classical approximation
captures the main features, but our LQC approximation calculation is capable 
of reducing the discrepancy substantially. Only at the latest times
this difference exceeds the level of the statistical errors. 
In Fig.~\ref{figN08b01} and Fig.~\ref{figN32b01} we 
display the corresponding results for  $N=8$ and $N=32$ with a sample 
of size $\mathcal{M}=4\times 10^{7}$.
Here we do not have an exact result to compare with, so the main issue
is the dependence of the errors on $N$ for a fixed sample size.
Errors of our method are larger than those of the classical
approximation as expected, but do not seem to depend crucially on the
number of variables.
The intermediate values of $N$, not shown, display  exactly the same
pattern.  For all values of $N$ the relative  difference between 
classical and linear QC approximation approaches a constant with 
errors  diminishing as the sample  size $\mathcal{M}$ grows.
For the maximum values studied of order  $\mathcal{M}=5 \times 10^{7}$,  
we can  estimate the  quantum correction at our latest times $mt\approx 3$ with
an accuracy of 10\% without a significant dependence on the number of 
degrees of freedom.

In conclusion, the proposed method seems to scale reasonably well with
the number of degrees of freedom. The computational cost is only a
certain factor higher than that involved in the classical
approximation. Thus, phenomenologically interesting applications are
addressable within present high performance computing capabilities. 

\section{Conclusions and Outlook}
\label{s.conclusions}
In this paper we have analyzed the real-time evolution  of simple
quantum systems. Both the form of the potential, as well as the type 
of initial conditions are chosen to reflect relevant applications 
in Cosmology. The simplicity of the systems allows the numerical
integration of the Schroedinger equation  and, hence, can be used to test different 
approximation methods. The simplest one is given by the 
{\em classical approximation} which amounts to  the classical evolution 
of a random variable distributed according to the initial
wave-function. In our examples the classical approximation always
performed well at initial times, capturing the qualitative features of
the quantum evolution.
When trying to improve on this approximation it seems
natural to focus on the Wigner function and the quantum Liouville 
evolution equation that it satisfies. It is simple to introduce a 
parameter $\kappa$ in the evolution equation that interpolates
between the classical approximation ($\kappa=0$) and the full-quantum 
evolution ($\kappa=1$). This parameter amounts to a rescaling in 
the value of $\hbar^2$ appearing explicitly in the equation. 

We have presented a method based on samples and a discretization 
(coarse-graining) in the distribution in the conjugate momentum, and 
compared its results with the classical approximation, the
next-to-leading 1/N  truncation of the  2PI equations,  and
the  numerical quantum evolution. The method can be used to determine 
the quantum corrections to the time-evolution of expectation values
as a power series expansion in $\kappa$. This provides a natural extension of 
the classical approximation (corresponding to the lowest-order). 
The results, for the examples considered,  seem to capture a sizable part
of the quantum effects, therefore providing a possible alternative to
other approaches. Even when the method sizably departed from the exact
quantum result, the discrepancy remained  smaller and of the order 
of the quantum effect, and no anomalous instabilities were observed. 
Thus, it can at least serve to attach a level of precision to the
classical approximation.

The method is certainly not a panacea. We emphasize that quantum
averages are subject to the {\em sign problem} since the Wigner
function is not positive definite. Our sampling method is equivalent 
to {\em reweighting}, which is certainly not a solution to the sign
problem. However, if one starts with a positive definite distribution
function, the problem only sets in at a later time, and reliable
estimates of the quantum effects can be obtained at early stages of
the evolution. The method can also be used to compute the full quantum
effect by setting $\kappa=1$. However at fixed $\kappa$, our coarse-grained
sampling method severely breaks down beyond a critical range of times.
However, one can go well beyond this point if one aims at computing the quantum 
corrections to order $\kappa^n$ and not the full quantum evolution.
In this case, the  computational cost only grows in a power-like fashion with respect to 
the time range of the analysis. This can then be  viewed as a systematic
improvement with respect to the classical approximation, which at the 
least  can serve to quantify the size of the quantum corrections
involved. All other errors, arising from the statistical size of the
sample or from the discretization in conjugate momenta are
quantifiable. 

As emphasized in the introduction, our main goal is to be able to apply
the method to the evolution of quantum fields in the early universe.
For that purpose it is important to see how the computational
cost depends on the number of degrees of freedom. Methods based on
histogramming the Wigner function have costs that grow exponentially
with the number of variables. In designing the methodology we focused 
on an approach which only grows in a power-like fashion, even if there
are more efficient ways to handle systems with a few degrees of
freedom. In the last section we have presented a pilot study to measure 
the rate at which the computational cost evolves with the number of
variables. The comparison is done by monitoring the size of the sample 
in order to keep the degree of accuracy of the quantum corrections fixed. 
In our exploratory study  we focused upon  a two-dimensional quantum
field theory case which has been subject of previous study as a toy
model for cosmological applications. The model has been discretized 
to a system with $N$ degrees of freedom. We applied our technique to 
the model up to $N=32$, and we found that the statistical errors on 
the quantum effects of our observables for a fixed sample size 
remain stable with the number of degrees of freedom. Of course, the
simulation time grows with the number of degrees of freedom, as does
for the case of the classical approximation. Even if computationally
demanding, these calculations are feasible with present day technology, 
and so will be the case for our proposed method. The results presented
in this paper allow us to be optimistic enough to embark in a higher 
scale study  with all the complexities involved in a quantum field 
theoretical treatment.

\begin{acknowledgments}
We thank Margarita Garc\'{\i}a P\'erez for a critical reading of the
manuscript and useful suggestions. 
We acknowledge financial support from the MCINN grants FPA2009-08785
and FPA2009-09017, the Comunidad Aut\'onoma
de Madrid under the program  HEPHACOS S2009/ESP-1473, and the European
Union under Grant Agreement number PITN-GA-2009-238353 (ITN STRONGnet).
The authors participate in the Consolider-Ingenio 2010 CPAN \linebreak
(CSD2007-00042). We acknowledge the use of the IFT clusters for part
of our numerical results.
\end{acknowledgments}


%

\end{document}